\pdfoutput=1
\documentclass[conference]{IEEEtran}
\usepackage[T1]{fontenc}
\usepackage[utf8]{inputenc}
\usepackage[english]{babel}
\usepackage[a4paper, total={184mm,239mm}]{geometry}
\usepackage{cite}
\usepackage{csquotes}
\usepackage{mathtools,amssymb,amsthm}
\usepackage{nicefrac}
\usepackage[binary-units=true]{siunitx}
\usepackage{tikz}
\usepackage{pgfplots}
\usepackage[textsize=tiny]{todonotes}

\makeatletter
\let\MYcaption\@makecaption
\usepackage{subcaption}
\let\@makecaption\MYcaption
\makeatother

\usepackage{booktabs}
\usepackage{microtype}
\usepackage[bookmarks=false,hidelinks]{hyperref}

\pgfplotsset{compat=1.15}
\usetikzlibrary{arrows.meta, backgrounds, calc, chains,shadows, matrix,positioning,shapes,shapes.geometric,decorations,shapes.arrows, decorations.pathmorphing, decorations.pathreplacing, decorations.markings, fit}

\newtheorem{example}{Example}
\newtheorem{lemma}{Lemma}
\newtheorem{definition}{Definition}

\newcommand{\ket}[1]{\ensuremath{|#1\rangle}}
\renewcommand{\baselinestretch}{0.9}

\hypersetup{
	pdftitle={As Accurate as Needed, as Efficient as Possible: Approximations in DD-based Quantum Circuit Simulation},
	pdfsubject={Design, Automation, and Test in Europe 2021},
	pdfauthor={Stefan Hillmich, Richard Kueng, Igor L. Markov, and Robert Wille}
}
\begin{document}
\date{}
\author{\IEEEauthorblockN{Stefan Hillmich\IEEEauthorrefmark{1}, Richard Kueng\IEEEauthorrefmark{1}, Igor L. Markov\IEEEauthorrefmark{2}, and Robert Wille\IEEEauthorrefmark{1}\IEEEauthorrefmark{3}}
\IEEEauthorblockA{\IEEEauthorrefmark{1}Institute for Integrated Circuits, Johannes Kepler University Linz, Austria}
\IEEEauthorblockA{\IEEEauthorrefmark{2}Department of EECS, University of Michigan, USA}
\IEEEauthorblockA{\IEEEauthorrefmark{3}Software Competence Center Hagenberg GmbH (SCCH), Austria}
{stefan.hillmich@jku.at, richard.kueng@jku.at, imarkov@eecs.umich.edu, robert.wille@jku.at}\\
{\small\url{https://iic.jku.at/eda/research/quantum/}}\vspace*{-1.4em}}
\title{\huge As Accurate as Needed, as Efficient as Possible: \\Approximations in DD-based Quantum Circuit Simulation\vspace{-.4cm}}	
\maketitle

\begin{abstract}
	Quantum computers promise to solve important problems faster than conventional computers. 
	However, unleashing this power has been challenging. 
	In particular, design automation runs into (1) the probabilistic nature of quantum computation and (2) exponential requirements for computational resources on non-quantum hardware.
	In quantum circuit simulation, \emph{Decision Diagrams} (DDs) have previously shown to reduce the required memory in many important cases by exploiting redundancies in the quantum state. 
	In this paper, we show that this reduction can be amplified by exploiting the probabilistic nature of quantum computers to achieve even more compact representations. 
	Specifically, we propose two new DD-based simulation strategies that approximate the quantum states to attain more compact representations, while, at the same time, allowing the user to control the resulting degradation in accuracy. 
	We also analytically prove the effect of multiple approximations on the attained accuracy and empirically show that the resulting simulation scheme enables speed-ups up to several orders of magnitudes.
\end{abstract}

\section{Introduction}
\label{sec:intro}

Quantum computing promises significant speed-ups for solving many important computational problems.
Oft-cited examples include Shor's algorithm~\cite{Sho:94} for integer factorization and Grover's search~\cite{Gro:96} for searching in unstructured databases.
However, other areas such as chemistry, finance, and machine learning can benefit from quantum computing as well~\cite{montanaro2016quantum,preskill2018quantum,coles2018quantum}.
This advantage largely comes from the exploitation of quantum-mechanical effects such as \emph{superposition} and \emph{entanglement}, where an \(n\)-qubit state can represent the \(2^n\) basis states at the same time and operations on one qubit can influence another qubit, respectively.
Those advantages have motivated Google, IBM, Microsoft, and Intel to build quantum chips and develop design tools, while start-ups like Cambridge Quantum and Rigetti have also joined the race.

Current laboratory and commercial quantum computers are exceedingly expensive, so most of the design and validation work, e.g.,~simulation~\cite{qiskit,jones2019quest,villalongaFlexibleHighperformanceSimulator2019}, compilation~\cite{DBLP:journals/tcad/AmyMMR13,sete2016functional,zulehnerEfficientMethodologyMapping2019}, and verification~\cite{brakerskiCryptographicTestQuantumness2018,gheorghiuVerificationQuantumComputation2019,burgholzerAdvancedEquivalenceChecking2020,burgholzerVerifyingResultsIBM2020}, is performed on non-quantum hardware.
Here, the power of quantum computing complicates design-tool development, since conventionally representing a quantum state requires an exponential amount of memory---often in the form of a vector with \(2^n\) entries representing the corresponding \emph{amplitudes} for each possible basis state.
Moreover, the straightforward representation of quantum operations is even worse with a \(2^n\!\times2^n\)-dimensional matrix.
This problem can be mitigated in many cases by employing adaptive data structures, such as \emph{Decision Diagrams} (DDs)~\cite{DBLP:books/daglib/0027785,DBLP:journals/tcad/ZulehnerW19,AP:06,WLTK:2008,DBLP:journals/tcad/NiemannWMTD16,DBLP:conf/iccad/ZulehnerHW19}, but the worst-case complexity remains exponential in the number of qubits.

Another important quantum-mechanical effect leads to further differences compared to non-quantum computing:
It is fundamentally impossible to observe the entire quantum state (i.e., the amplitudes for all basis states) without destroying the superposition and entanglement~\cite{NC:2000}.
Instead, a measurement of a quantum state collapses the state into one of the possible basis states.
Since such a measurement is probabilistic (and depends on the respective amplitudes for each basis state), the quantum computation as a whole is probabilistic as well, which complicates applications.
And yet, our work shows how to improve design-tool efficiency using the probabilistic nature of quantum computation. 

Small changes in the amplitudes of a quantum state lead to small changes in the probabilities of measurement outcomes. 
Accordingly, we can manipulate the quantum state to admit a more compact representation in a finely-controlled tradeoff with its accuracy.
In other words, the probabilistic nature of quantum computation makes, to some degree, quantum computations resistant against such small manipulations.
Depending on the simulated quantum algorithm and/or specific circuit, a low-accuracy approximation of the final state may still be suitable for non-quantum post-processing leading to the same results, perhaps, after several repeated attempts.
For example, Shor's algorithm works reliably with a circuit fidelity of \SI{50}{\percent} (as we show in Section~\ref{sec:evaluation}), while the quantum-supremacy circuits from Google still provide meaningful results with circuit fidelity around~\SI{1}{\percent}~\cite{arute2019quantum,markov2020massively}.
This is a truly novel feature of quantum computations compared to computations in the conventional domain which has hardly been exploited yet.

In this work, we propose to exploit error tolerance to speed up DD-based quantum circuit simulation on non-quantum hardware.
We investigate applying multiple approximation rounds throughout the simulation process to attain a more compact decision diagram representing the quantum state and, thus, improve the simulation runtime.
To this end, we present two dedicated methods to incorporate approximation into the simulation process.
The first method is reminiscent of automatic \emph{garbage collection} used with programming languages and triggers the approximation when the intermediate quantum state grows too large in terms of size of the decision diagram.
The second method facilitates the simulation with arbitrary approximation as long as a certain minimum fidelity in the resulting quantum state is guaranteed. 
Empirical validation demonstrates speed-ups in quantum circuit simulation up to several orders of magnitude.

The remainder of this paper is structured as follows:
Section~\ref{sec:background} reviews the fundamentals of quantum computations and decision diagrams.
In Section~\ref{sec:motivation}, we describe key observations that motivate the proposed methods.
In Sections~\ref{sec:approximation-simulation} and~\ref{sec:theory}, we propose two methods for approximations in DD-based quantum circuit simulation and provide a proof on the effect of multiple approximations on the overall fidelity, respectively.
Section~\ref{sec:evaluation} summarizes the empirical validation of the proposed methods.
Finally, Section~\ref{sec:conclusions} concludes the paper.

\section{Background}
\label{sec:background}

To keep the paper self-contained, this section provides a brief overview of the basics of quantum computing and decision diagrams as a means to represent quantum states.

\subsection{Quantum Computing}
\label{sec:qc-background}

The basic unit of information in quantum computing is the \emph{quantum bit} or \emph{qubit}~\cite{NC:2000}.
In the conventional realm, a bit can assume exactly one of the states \(0\) and \(1\).
Qubits can additionally assume any linear combination of the \emph{basis states} (denoted \ket{0} and \ket{1} in Dirac notation).
More precisely, the state of a qubit is written as \( \ket{\psi} = \alpha_0\cdot\ket{0} + \alpha_1\cdot\ket{1} \) with \emph{amplitudes} \(\alpha_0, \alpha_1 \in \mathbb{C} \).
The squared magnitude \(|\alpha_i|^2\) of an amplitude \(\alpha_i\) defines the probability with which the corresponding basis state will be the result when measuring.
Therefore, the amplitudes have to satisfy the normalization constraint \( |\alpha_0|^2 + |\alpha_1|^2 = 1 \).
A qubit is said to be in \emph{superposition}, when both \(\alpha_0\) and \(\alpha_1\) are \mbox{non-zero}. 
Intuitively, this means the qubits are in both states at the same time---one of the important characteristics of quantum computing.
Another important characteristic is \emph{entanglement}, where the measurement of a single qubit may influence the (future) measurement result of another qubit.

Due to the underlying physics, the exact values of the amplitudes (\(\alpha_0\) and \(\alpha_1\)) are fundamentally unobservable in physical quantum computers.
Instead, the only way to \enquote{look at} qubits is measuring (with the outcome probabilities dictated by the amplitudes).
Measuring a qubit \( \ket{\psi} = \alpha_0\cdot\ket{0} + \alpha_1\cdot\ket{1} \) will yield \(\ket{0}\) (\ket{1}) with the probability ~\( |\alpha_0|^2 \) (\( |\alpha_1|^2 \)).
Further, the measurement destroys any superposition and entanglement---leaving the qubit in a basis state.

Quantum states consisting of \(n\) qubits are extended accordingly to have \(2^n\) basis states and corresponding amplitudes.
The normalizing constraint is generalized to \(\sum_{i\in \{0,1\}^n} |\alpha_i|^2 = 1\).
Further, quantum states are commonly described by vectors with the amplitudes as elements, e.g.,~a two-qubit state \(\ket{\psi}\) may be denoted as~\(\begin{bsmallmatrix}\alpha_{00}& \alpha_{01}& \alpha_{10}& \alpha_{11}\end{bsmallmatrix}^\mathrm{T}  \).
\begin{example}
	\label{ex:qubits}
	Consider a two-qubit quantum state \(\ket{\psi}\), which is set to \mbox{\(\nicefrac{1}{\sqrt{2}}\cdot\ket{00} + 0\cdot\ket{01} + 0\cdot\ket{10} +\nicefrac{1}{\sqrt{2}}\cdot\ket{11}\)}. 
	This state is valid, since \({\left|\nicefrac{1}{\sqrt{2}}\right|^2 + \left|\nicefrac{1}{\sqrt{2}}\right|^2 = 1}\) satisfies the normalizing constraint. 
	As a vector, the state is written as
	\(
		\ket{\psi} = \begin{bsmallmatrix} \nicefrac{1}{\sqrt{2}}& 0& 0& \nicefrac{1}{\sqrt{2}} \end{bsmallmatrix}^\mathrm{T}.
	\)
	Due to the superposition, measuring this state yields either of the two basis states \(\ket{00}\) or \(\ket{11}\) with a probability of \(\left| \nicefrac{1}{\sqrt{2}}\right|^2 = \nicefrac{1}{2}\) each.
	After the measurement, the superposition is destroyed and the quantum state is fixed to the measured state, i.e., subsequent measurements  yield the same result.
\end{example}
A quantum state is altered quantum operations.
These operations are defined through unitary matrices, i.e.,~square matrices whose inverse is their conjugate transpose~\cite{NC:2000}.
\begin{example}
	Commonly used single-qubit operations are ${X}$ (negates the state of the qubit), ${H}$ (sets the qubit into superposition), and ${Z}$ (shifts the phase of the qubit).
	
	A key example for a two-qubit operation is the \(\mathit{CNOT}\), which negates a target qubit iff the control qubit is in the state \ket{1}.
\end{example}

Given the vector- and matrix-based descriptions for states and operations, respectively, the effect of applying an operation to a state is defined through the matrix-vector multiplication.
\begin{example}\label{ex:qua_op}
	Given a quantum state \(\ket{\psi}\) with two qubits set to~\(\ket{00}\), first	performing a Hadamard operation on the first qubit and, afterwards, a \(\mathit{CNOT}\) operation yields a new state \(\ket{\psi'}\) defined by 
	\[
		\underbrace{\begin{bsmallmatrix} 1 & 0 & 0 & 0 \\ 0 & 1 & 0 & 0 \\ 0 & 0 & 0 & 1 \\ 0 & 0 & 1 & 0 \end{bsmallmatrix}}_{\mathit{CNOT}} 
		\,\times 
		\underbrace{\frac{1}{\sqrt{2}}\begin{bsmallmatrix*}[r]
			1 & 0 & 1 & 0 \\
			0 & 1 & 0 & 1 \\
			1 & 0 & -1 & 0 \\
			0 & 1 & 0 & -1 \\
		\end{bsmallmatrix*}}_{\text{Hadamard on 1st qubit}}
		\times
		\underbrace{\begin{bsmallmatrix} 1 \\ 0 \\ 0 \\ 0 \end{bsmallmatrix}}_{\ket{\psi}}
		= 
		\underbrace{\frac{1}{\sqrt{2}}\begin{bsmallmatrix} 1 \\ 0 \\ 0 \\ 1 \end{bsmallmatrix}}_{\ket{\psi'}}.
	\]
	Possible outcomes in measuring the new state are \(\ket{00}\) or \(\ket{11}\)---each with a probability of \SI{50}{\percent} as in Example~\ref{ex:qubits}.
\end{example}

\subsection{Decision Diagrams} \label{sec:dds}

Na\"ive representations of quantum states and operations rely on exponentially large vectors and matrices of size~$2^n$ and $2^n\!\times 2^n$, respectively (with $n$ denoting the number of qubits).  
\emph{Decision diagrams}~(DDs) are tried and tested alternative representations that exploit redundancies in order to reduce these complexities~\cite{DBLP:conf/iccad/BaharFGHMPS93,DBLP:books/daglib/0027785,DBLP:journals/tcad/ZulehnerW19,AP:06,WLTK:2008,DBLP:journals/tcad/NiemannWMTD16}.
In the task of simulation, utilizing decision diagrams instead of vectors/matrices increases the performance by several orders of magnitude for certain algorithms~\cite{DBLP:journals/tcad/ZulehnerW19}.
In the best case, this resulted in a runtime of under two minutes with decision diagrams compared to 30 days with matrix-vector multiplication.

Redundancies are exploited by using shared structures whenever possible.
To this end, e.g., the vector is split into upper and lower sub-vectors.
This splitting is repeated for the sub-vectors until the result is a single element.
Since each split halves the size of the vector, there are \(n\) levels of splitting for an \(n\)-qubit state.
During this process, identical sub-vectors are detected and represented by a single shared structure for the identical sub-vectors. 
The decision diagram is then normalized to ensure a canonical representation.
To determine the value of an amplitude, the edge weights of the path representing said amplitude are multiplied.
Analogously, this is done for matrices.
	
\begin{figure}[tbp]
	\centering
	\begin{subfigure}[t]{0.25\linewidth}
		\centering
		\scalebox{0.9}{\begin{tikzpicture}
		\matrix[matrix of math nodes, left delimiter={[},right delimiter={]}, inner xsep=0] (vector) {
			\nicefrac{1}{\sqrt{10}}\\
			0\\
			0\\
			\nicefrac{-1}{\sqrt{10}}\\
			0\\
			\nicefrac{2}{\sqrt{10}}\\
			0\\
			\nicefrac{2}{\sqrt{10}}\\
		};
		
		\begin{scope}[on background layer, gray]	
		\node[right=-0.6cm of vector-1-1.center, anchor=east] {\(\ket{000}\)};
		\node[right=-0.6cm of vector-2-1.center, anchor=east] {\(\ket{001}\)};
		\node[right=-0.6cm of vector-3-1.center, anchor=east] {\(\ket{010}\)};
		\node[right=-0.6cm of vector-4-1.center, anchor=east] {\(\ket{011}\)};
		\node[right=-0.6cm of vector-5-1.center, anchor=east] {\(\ket{100}\)};
		\node[right=-0.6cm of vector-6-1.center, anchor=east] {\(\ket{101}\)};
		\node[right=-0.6cm of vector-7-1.center, anchor=east] {\(\ket{110}\)};
		\node[right=-0.6cm of vector-8-1.center, anchor=east] {\(\ket{111}\)};
		\end{scope}
		\end{tikzpicture}}
		\caption{Vector}
		\label{fig:statevectorvector}
	\end{subfigure}
	\quad
	\begin{subfigure}[t]{0.25\linewidth}
		\centering
		\scalebox{0.9}{\begin{tikzpicture}[terminal/.style={draw,rectangle,inner sep=0pt}]	
		\matrix[matrix of nodes,ampersand replacement=\&,every node/.style={draw,circle,inner sep=0pt,minimum width=0.5cm,minimum height=0.5cm},column sep={1cm,between origins},row sep={1cm,between origins}] (qmdd2) {
										    \& |(m1)|$q_2$ \\
			|[xshift=+0.5cm](m2b)| $q_1$	\&                                  \& |[xshift=-1em](m2a)| $q_1$ \\
			|(m3b)| $q_0$                   \& |[xshift=-0.00cm](m3c)| $q_0$ 	\& |[xshift=-1em](m3a)| $q_0$ \\
			                                \& |[terminal] (t3)| $1$ \\
		};
		
		\draw[very thick] ($(m1)+(0,0.7cm)$) -- (m1) node[right, midway]{$\frac{2}{\sqrt{10}}$};
		
		\draw (m1) -- ++(300:0.6cm) -- (m2a);
		\draw[very thick] (m1) -- ++(240:0.6cm) node[left, midway] {$\frac{1}{2}$} -- (m2b);
		
		\draw (m2a) -- ++(240:0.6cm) -- (m3a);
		\draw (m2a) -- ++(300:0.6cm) -- (m3a);
		
		\draw (m2b) -- ++(240:0.6cm) -- (m3b);
		\draw[very thick] (m2b) -- ++(300:0.6cm) node[right, midway] {$-1$} -- (m3c);
		
		\draw (m3a) -- ++(240:0.4cm) node[below, xshift=0.5pt, inner sep=0,font=\scriptsize] {$0$};
		\draw (m3a) -- ++(300:0.6cm) -- (t3);
		
		\draw (m3b) -- ++(240:0.6cm) -- (t3);
		\draw (m3b) -- ++(300:0.4cm) node[below, xshift=0.5pt, inner sep=0,font=\scriptsize] {$0$};
		
		\draw (m3c) -- ++(240:0.4cm) node[below, xshift=0.5pt, inner sep=0,font=\scriptsize] {$0$};
		\draw[very thick] (m3c) -- ++(300:0.6cm) -- (t3);
		\end{tikzpicture}}
		\caption{DD}
		\label{fig:dd-statevector}
	\end{subfigure}
	\qquad
	\begin{subfigure}[t]{0.125\linewidth}
		\centering
		\scalebox{0.9}{\begin{tikzpicture}
			\matrix[matrix of math nodes, left delimiter={[},right delimiter={]}, inner xsep=0] (vector) {
				0\\
				0\\
				0\\
				0\\
				0\\
				\nicefrac{1}{\sqrt{2}}\\
				0\\
				\nicefrac{1}{\sqrt{2}}\\
			};
		\end{tikzpicture}}
		\caption{Vector}
		\label{fig:dd-statevector-approximated-vector}
	\end{subfigure}
	\begin{subfigure}[t]{0.125\linewidth}
		\centering
		\scalebox{0.9}{\begin{tikzpicture}[terminal/.style={draw,rectangle,inner sep=0pt}]	
		\matrix[matrix of nodes,ampersand replacement=\&,every node/.style={draw,circle,inner sep=0pt,minimum width=0.5cm,minimum height=0.5cm},column sep={1cm,between origins},row sep={1cm,between origins}] (qmdd2) {
			\& |(m1)|$q_0$ 											\\
			\& |(m2a)| $q_1$                                 \&  						\\
			\& |(m3a)| $q_2$ 					\&  	\\
			\& |[terminal] (t3)| $1$ 									\\
		};
		
		\draw ($(m1)+(0,0.7cm)$) -- (m1) node[right, midway]{$\frac{1}{\sqrt{2}}$};
		
		\draw (m1) -- ++(300:0.6cm) -- (m2a);
		\draw (m1) -- ++(240:0.4cm) node[below, xshift=0.5pt, inner sep=0,font=\scriptsize] {$0$};
		
		\draw (m2a) -- ++(240:0.6cm) -- (m3a);
		\draw (m2a) -- ++(300:0.6cm) -- (m3a);
		
		\draw (m3a) -- ++(240:0.4cm) node[below, xshift=0.5pt, inner sep=0,font=\scriptsize] {$0$};
		\draw (m3a) -- ++(300:0.6cm) -- (t3);
		\end{tikzpicture}}
		\caption{DD}
		\label{fig:dd-statevector-approximated-dd}
	\end{subfigure}
	\caption{Two quantum states with vector and DD representation, respectively}
	\label{fig:bg-state}
	\vspace{-1em}
\end{figure}
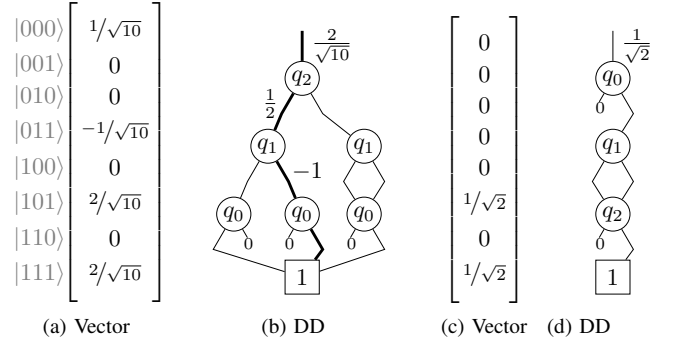

\begin{example} 
	Consider the state vector in Fig.~\ref{fig:statevectorvector}. 
	The annotations on the left denote the basis state each amplitude corresponds to.
	In Fig.~\ref{fig:dd-statevector}, a decision diagram representing the same state is depicted.
	To access the amplitude of basis state \ket{011}, the bolded path in the decision diagram has to be traversed and the edge weights along this path have to be multiplied, i.e., (\(q_2 = 0\), \(q_1 = 1\), \(q_0 = 1\)) yielding  \(\nicefrac{2}{\sqrt{10}} \cdot \nicefrac{1}{2} \cdot (-1) \cdot 1 = \nicefrac{-1}{\sqrt{10}} \).
\end{example}

\section{Motivation}
\label{sec:motivation}

In this work, we investigate how to efficiently simulate quantum computations on non-quantum machines.
Computational efficiency aside, this task is straightforward to solve and boils down to a series of matrix-vector multiplications as reviewed in Section~\ref{sec:qc-background}. 
However, this approach requires an exponential amount of memory to store the corresponding vectors and matrices.
Decision diagrams as reviewed in Section~\ref{sec:dds} may cope with that for many instances, but eventually have the same worst-case complexity.
We propose to improve upon that state of the art by exploiting approximations of the respectively considered decision diagrams.

Physical quantum computers are inherently probabilistic in nature, i.e., they do not output an explicit vector of amplitudes but rather a bitstring representing a single basis state which is probabilistically selected during measurement. 
While the results of such measurements still depend on the respective amplitudes, \SI{100}{\percent} accuracy in the amplitudes is not always necessary for that.
For example, near-zero amplitudes imply \mbox{near-zero} probabilities and, thus, may be ignored in many quantum circuit applications without much impact on the final result. 
In this work, we are going to exploit this observation. 
To this end, we first require a metric to quantify proximity for quantum states.

The \emph{fidelity metric} of two quantum states measures their proximity~\cite{NC:2000,watrous2018theory}.
It expresses the likelihood that measuring two quantum states results in the same outcome. 
\begin{definition}[Fidelity]\label{def:fidelity}
	For pure quantum states as used in this work, the \emph{fidelity} \(F\) is defined as \( F(\ket{\psi}, \ket{\phi}) = |\langle\psi|\phi\rangle|^2 = |\left(\psi^*\right)^\mathrm{T} \cdot \phi|^2\)
	for any quantum states \(\ket{\psi}\) and \(\ket{\phi}\)~\cite{watrous2018theory}.
	Further, \(0 \le F(\ket{\psi}, \ket{\phi}) \le 1\) holds, where \(F(\ket{\psi}, \ket{\phi}) = 1\) implies \(\ket{\psi} = \ket{\phi}\).
\end{definition}

\begin{example}\label{ex:fidelity}
	Given the two quantum states \(\ket{\psi} = \nicefrac{1}{2}\begin{bsmallmatrix}1 & 1 & 1 & 1\end{bsmallmatrix}^\mathrm{T}\) and \(\ket{\phi} = \nicefrac{1}{\sqrt{2}}\begin{bsmallmatrix}1 & 0 & 0 & 1\end{bsmallmatrix}^\mathrm{T}\), their fidelity is calculated as	
	\(
	F(\ket{\psi}, \ket{\phi}) = \left|\nicefrac{1}{2}\cdot\nicefrac{1}{\sqrt{2}} + \nicefrac{1}{2}\cdot\nicefrac{1}{\sqrt{2}}\right|^2 = \nicefrac{1}{2}.
	\)
	Intuitively, this states that the probability of measuring the same outcome from both quantum states is \SI{50}{\percent} (which indeed is the case for these two quantum states).
\end{example}

We use the fidelity metric to determine the accuracy of approximation, i.e.,~how far the original state is from its approximation obtained by zeroing out some of its small amplitudes.
In this context, the fidelity metric has an important property: it is preserved under unitary transformations, i.e.,~the quantum operations we use in simulation.
More precisely, given a quantum operation \(U\) and two quantum states~\(\ket{\psi}\)~and~\(\ket{\phi}\), applying \(U\) to each state does not further decrease their fidelity, i.e., \(F(U\ket{\psi}, U\ket{\phi}) = F(\ket{\psi}, \ket{\phi})\)~\cite{watrous2018theory}. 
Therefore, small amplitudes can be zeroed out during the simulation while maintaining a reasonable end-to-end accuracy.

Furthermore, even though zeroing out small amplitudes a single time during simulation might be sufficient, multiple applications may produce better size-fidelity tradeoffs---in particular, when larger instances of quantum algorithms are considered. 
This raises the problem of how to calculate the fidelity between the exact state (where no amplitudes were ignored) and the modified state (where amplitudes were ignored multiple times during simulation).

\begin{example}\label{ex:multi-approx}
	Given the original state \mbox{\(\ket{\psi} = \nicefrac{1}{2}\begin{bsmallmatrix}1 & 1 & 1 & 1\end{bsmallmatrix}^\mathrm{T}\)}, we are generating two states with successively more amplitudes set to zero (as an exaggerated example).
	We obtain \mbox{\(\ket{\psi'} = \nicefrac{1}{\sqrt{2}}\begin{bsmallmatrix}1 & 0 & 0 & 1\end{bsmallmatrix}^\mathrm{T}\)} in the first round and \(\ket{\psi''} = \begin{bsmallmatrix}0 & 0 & 0 & 1\end{bsmallmatrix}^\mathrm{T}\) in the second.
	The pairwise fidelities are 
		\(F(\ket{\psi}, \ket{\psi'}) = \nicefrac{1}{2}\), \(F(\ket{\psi'}, \ket{\psi''}) = \nicefrac{1}{2}\), and \(F(\ket{\psi}, \ket{\psi''}) = \nicefrac{1}{4}\).
\end{example}

The proposed approximation method of zeroing out small amplitudes and rescaling the vector constitutes a worst case for the fidelity metric.
In Section~\ref{sec:theory}, we back up the intuition that successive applications lower the overall fidelity multiplicatively by a formal statement with a proof.
This way we can repeatedly decrease the size of the representation of a quantum state during simulation while, at the same time, being able to precisely keep track of the resulting accuracy.

\section{Approximating DD-based \\Quantum Circuit Simulation}
\label{sec:approximation-simulation}

The discussions in Section~\ref{sec:motivation} eventually provide the basis for a \mbox{DD-based} quantum circuit simulation approach which allows the user to simplify (to \emph{approximate}) decision diagrams (and, by that, to accelerate the simulation process), while controlling the resulting accuracy.
In this section, we describe a corresponding implementation of such an approach.
To this end, we first review how approximation of decision diagrams can actually be employed. 
Afterwards, two strategies that incorporate the approximation into the simulation process are presented.

\subsection{Constructively Approximating DDs}\label{sec:approx}

Before introducing our approximation techniques, we briefly review how decision diagrams are used in simulation.
Performing quantum circuit simulations based on decision diagrams is conceptionally similar to the matrix-vector-based approach discussed in Section~\ref{sec:qc-background}.
Decision diagrams can represent quantum operations and states just like matrices and vectors, respectively.
Additionally, decision diagrams support the same operations, especially matix-vector multiplication, 
which are required for circuit simulation~\cite{DBLP:books/daglib/0027785,DBLP:journals/tcad/ZulehnerW19,DBLP:journals/tcad/NiemannWMTD16,zulehner2019matrix}.
Hence, on an abstract level, a decision diagram representing a basis state is constructed and, afterwards, the quantum operations (again, represented by decision diagrams) are successively applied to the quantum state.

A method to efficiently employ approximation as motivated in the previous section 
must be able to 
(1)~remove nodes from the decision diagrams while, at the same time,
(2)~efficiently determine the effect of that removal on the accuracy (i.e., the fidelity).
Moreover, a corresponding method should ideally guarantee that the accuracy/fidelity has a lower bound, which can be defined by the user.
These objectives can be accomplished by numerically estimating the \emph{norm contribution} of each node to the overall accuracy and selecting appropriate nodes for removal~\cite{DBLP:conf/aspdac/ZulehnerHMW20}.

\begin{definition}[Norm Contribution of a Node]
	The paths through a decision diagram from top to bottom encode the amplitudes in the edge weights. 
	More precisely, the product of edge weights along a path yields the corresponding amplitude.
	The \emph{norm contribution of a node} (or \emph{contribution} for short) is the sum of squared magnitudes of amplitudes for each path passing through that node.
	From this, it follows that for each level \(i\) in the decision diagrams, the contributions of nodes \(q_i\) on this level add up to 1.
\end{definition}

\begin{example}
	Consider again the decision diagram shown in Fig.~\ref{fig:dd-statevector}.
	Since all paths of the decision diagram go through the root node labeled \(q_2\), this node has a contribution of~1.
	The nodes labeled \(q_1\) and \(q_0\) on the right-hand side have a contribution of 0.8 each, because the squared magnitude of amplitudes for paths passing through is \(2\cdot|\nicefrac{2}{\sqrt{10}}|^2=0.8\).
	The node labeled~\(q_1\) on \mbox{left-hand} side has a contribution of 0.2 and its two \(q_0\)-successors have a contribution of 0.1 each. 
\end{example}
\vspace{200em}

Having calculated the contributions of each node, the effect of removing a node from the decision diagram on the resulting fidelity can be determined:
since the removal of a node corresponds to setting the relevant amplitudes to zero, the fidelity decreases additively by the sum of squared magnitude of the zeroed amplitudes.
Hence, the contribution of a node directly translates into the fidelity lost on removal. 
This facilitates a lower bound on the resulting fidelity by only removing nodes with a small enough contribution.

\begin{example}
	Consider again the decision diagram shown in Fig.~\ref{fig:dd-statevector} and the corresponding node contributions as determined above.
	Removing the root node labeled \(q_2\) (and, by this, the entire decision diagram) would lead to a fidelity of 0---accordingly reflected by the node's contribution of~1.
	Instead, removing the left-hand side node labeled \(q_1\) (which has a contribution of 0.2) results in the state shown in Fig.~\ref{fig:dd-statevector-approximated-dd} as discussed before---yielding a much more compact representation while maintaining a fidelity of 0.8.
\end{example}

The above technique can be used at any point during simulation to evaluate the effect of removing individual nodes in terms of implied accuracy loss.
Based on that, we can now focus on specific strategies to ensure good accuracy-efficiency tradeoffs. 
These are described next.

\subsection{Memory-driven Approximation} 
\label{sec:memory-driven}
First, we propose a reactive strategy which focuses on efficiency, i.e.,~caters to the use case, where the size of the decision diagram (and, hence, the required memory requirements) should be kept low---even if this means risking an unsuitable loss in fidelity.
To this end, we evaluate after each simulation step (i.e.,~after each application of a quantum operation to the current state) whether the size of the decision diagram (i.e.,~its number of nodes)
exceeds a certain threshold (defined by the user). 
This is similar to typical garbage collection methods in which memory is freed once certain thresholds are exceeded (the difference here is that we do this at the expense of accuracy). 
Since the fidelity decreases with each application of such an \emph{approximation round}, the threshold is dynamically increased as well to avoid that the number of applied approximations gets too large.

We distill the ideas proposed so far into the following procedure: in the simulation, the quantum operations are successively applied to the quantum state. 
After each quantum operation, the size of the decision diagram representing the state is compared to the user-defined threshold. If that threshold is exceeded, the quantum state is approximated targeting the \emph{single-round fidelity} (which is also given by the user).
Additionally, the threshold is doubled after each approximation round to avoid too many approximations.

Due to its simple structure, this strategy is appropriate as a reactive protection against out-of-memory conditions, especially when simulating a new circuit type and not knowing what to expect.
\begin{example}
	The quantum-supremacy circuits~\cite{boixo2016characterizing} are designed so that they posses little to no redundancy, which is an exceptionally challenging case for quantum state representation by decision diagrams.
	Therefore, the size of the decision diagram will increase rapidly to the point were it significantly slows down the simulation process.
	At that point, the approximation scheme proposed here kicks in and trades off some accuracy for a smaller representation and subsequently faster simulation.
	This process repeats when the decision diagram reaches the new threshold size.
\end{example}

Underestimating the hyper-parameters for the threshold and single-round fidelity may render the simulation result meaningless if the final state fidelity is too low.
Nonetheless, when carefully considering the algorithm at hand, memory-driven approximation may enable successful simulations previously blocked by insufficient memory.

\subsection{Fidelity-driven Approximation}

To complement the \emph{reactive} strategy from Section~\ref{sec:memory-driven}, we now propose a \emph{proactive} strategy based on accuracy. 
This strategy caters to the use case where the fidelity must not drop below a certain lower bound but, beyond that, the decision diagram can be approximated as much as possible.
Indeed, in many applications, it is sufficient if the resulting quantum state is not exact but employs a certain lower-bound accuracy.
For example, Shor's algorithm~\cite{Sho:94} frequently works with a circuit fidelity of around \SI{50}{\percent} and still determines the factors of an integer correctly as we show in Section~\ref{sec:evaluation}.
Since the approximation method described in Section~\ref{sec:approx}, i.e.,~the removal of nodes of a decision diagram, can guarantee a certain lower bound for fidelity, this can easily be realized. 

Given a minimum required final fidelity \(f_\text{final}\) for the quantum state after the simulation, the number of times the approximation is applied (\emph{approximation rounds}) is proactively calculated for the circuit.
Due to the multiplicative property of the fidelity metric, the maximum number of approximation rounds is \(\lfloor\log_{f_\text{round}}(f_\text{final})\rfloor\), where \(f_\text{round}\) is the target fidelity for each approximation round.
This requires choosing a tradeoff between (1) few approximation rounds with low single-round fidelity and (2) many approximation rounds with high single-round fidelity.
The first choice keeps the overhead of the approximation rounds low, but in between the size of the decision diagram may grow too much.
The second choice, on the other hand, is much more likely to limit the growth of the decision diagram while imposing the overhead of performing many approximation rounds.
The optimal selection heavily depends on the quantum algorithm.

In addition to deciding the number of approximation rounds, the locations in the quantum circuit at which to approximate have to be determined.
From a high-level perspective, promising candidates for such locations are between circuit blocks of the algorithm.
When no such circuit blocks can be identified, e.g.,~after certain types of circuit optimization, the individual approximation rounds are evenly spaced out through the circuit.

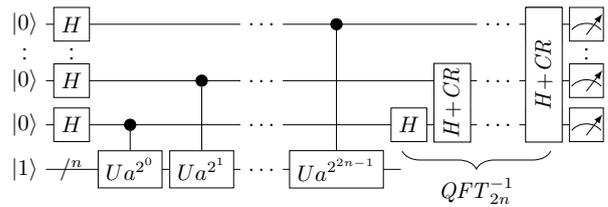
\begin{figure}
	\centering
	\resizebox{0.9\linewidth}{!}{\begin{tikzpicture}[gate/.style={draw,fill=white,minimum size=1.5em},
			control/.style={draw,fill,shape=circle,minimum size=5pt,inner sep=0pt},
			meter/.append style={
					draw,
					rectangle, 
					minimum size=1.5em, 
					fill=white,
					path picture={
						\draw[black] ([shift={(1pt,5pt)}]path picture bounding box.south west) to[bend left=50] ([shift={(-1pt,5pt)}]path picture bounding box.south east);
						\draw[black,-latex] ([shift={(0,.1)}]path picture bounding box.south) -- ([shift={(.25,-.1)}]path picture bounding box.north);}}]
			\matrix (table) [%
	     matrix of nodes,
	     nodes in empty cells,
	     ampersand replacement=\&,
	     row sep=3pt,
	     column sep=3pt,
	     row 1/.style={row sep=0pt, inner ysep=0pt, outer ysep=0pt},
	     row 2/.style={row sep=-0pt, inner ysep=-3pt, outer ysep=-0pt, nodes={}},
	     every node/.style={anchor=center},
	   ] {%
	     \ket{0}        \& |[gate]| \(H\) \&                       \&                       \& \(\ldots\) \& |[control]|                \&                \&                \&  \(\ldots\) \& |[gate]| \(H\)  \& |[meter]| \\
	     \({\vdots}\) \& \({\vdots}\)   \&                       \&                       \&            \&                            \&                \&                \& \&   \& \({\vdots}\)    \\
	     |[fill=white]|\ket{0}        \& |[gate]| \(H\) \&                       \& |[control]|           \& \(\ldots\) \&                            \&                \& |[gate]| \(H\) \&  \(\ldots\) \&   \& |[meter]| \\
	     \ket{0}        \& |[gate]| \(H\) \& |[control]|           \&                       \& \(\ldots\) \&                            \& |[gate]| \(H\) \& |[gate]| \(H\) \&  \(\ldots\) \& |[gate]| \(H\) \& |[meter]| \\
	     \ket{1}        \& {\(/^n\)}      \& |[gate]| \(Ua^{2^0}\) \& |[gate]| \(Ua^{2^1}\) \& \(\ldots\) \& |[gate]| \(Ua^{2^{2n-1}}\) \&  \& \\                           
	  };
	  
		 \node[fit=(table-3-8.north west)(table-4-8.south east), draw,fill=white, inner sep=0pt] (qft2) {};
		 \node[fit=(table-1-10.north west)(table-4-10.south east), draw,fill=white, inner sep=0pt] (qft3) {};
		 
		 \node [draw, rotate=90, draw=none] at (qft2) {\(\mathit{H{+}CR}\)};
		 \node [draw, rotate=90, draw=none] at (qft3) {\(\mathit{H{+}CR}\)};
		 
		 \draw [decorate,decoration={brace,amplitude=8pt,raise=4pt,mirror}] ($(table-4-7.south)-(3pt,0)$) -- ($(table-4-10.south)+(3pt,0)$) node [black,midway,below, yshift=-1.25em,align=center] {\(\mathit{QFT}_{2n}^{-1}\)};
		 
		 \begin{scope}[on background layer]
				\draw (table-1-1.east) -- (table-1-5.west);
				\draw (table-3-1.east) -- (table-3-5.west);
				\draw (table-4-1.east) -- (table-4-5.west);
				\draw (table-5-1.east) -- (table-5-5.west);
				
				\draw (table-1-5.east) -- (table-1-9.west);
				\draw (table-3-5.east) -- (table-3-9.west);
				\draw (table-4-5.east) -- (table-4-9.west);
				\draw (table-5-5.east) -- (table-5-7);
				
				\draw (table-1-9.east) -- (table-1-11.west);
				\draw (table-3-9.east) -- (table-3-11.west);
				\draw (table-4-9.east) -- (table-4-11.west);
				
				\draw (table-4-3) -- (table-5-3);
				\draw (table-3-4) -- (table-5-4);
				\draw (table-1-6) -- (table-5-6);
		 \end{scope}
	\end{tikzpicture}}
	\vspace{-1em}
	\caption{Circuit blocks of Shor's algorithm}
	\label{fig:shor}
	\vspace{-1.25em}
\end{figure}

\begin{example}
	Shor's algorithm~\cite{Sho:94} exhibits distinctive circuit blocks as illustrated in Fig.~\ref{fig:shor}.
	It consists of Hadamard operations, a series of modular multiplications (\(Ua^x\)), and an inverse QFT, which itself is further split into Hadamard operations and controlled rotations (\!\(\mathit{CR}\)).
	Exploiting this knowledge facilitates approximation rounds after each modular multiplication and after the controlled rotations during the inverse QFT.
	Following the proposed strategy, the number of approximation rounds and their positions are determined before the simulation to lower-bound the fidelity of the quantum state after the simulation.
\end{example}

As with the reactive strategy, suitable hyper-parameter selection is paramount, because producing results with unnecessarily high fidelity may require greater computational resources.

\vspace{1em}
\section{Effect of Multiple Approximations}
\label{sec:theory}

The effect of the applied approximations is crucial to the eventual accuracy and, hence, the approaches proposed above.
In the motivation in Section~\ref{sec:motivation} and the approaches proposed in Section~\ref{sec:approximation-simulation}, we established the contribution of nodes to directly calculate the effect on the fidelity if said nodes are removed.
Moreover, we claimed that even applying several approximation rounds can be reflected by multiplying the correspondingly resulting fidelities.
However, while this claim may seem intuitive, it lacks a rigorous underpinning thus far.
We now provide theoretical basis for this claim. 

The \emph{fidelity metric} quantifies the proximity of quantum states. 
Recall Definition~\ref{def:fidelity}: Given two quantum states $\ket{\psi}, \ket{\phi} \in \mathbb{C}^D$: \(F \left( \ket{\psi}, \ket{\phi} \right) = \left| \langle \psi | \phi \rangle \right|^2 \in \left[0,1\right]\) for \mbox{\(D=2^n\)-dimensional} states.
In the following, we show that the fidelity metric behaves as one might expect under successive approximations.
More precisely, we consider a simple \emph{truncation procedure}, which is a generalization of the proposed approximation scheme for decision diagrams.
Fix a subset \mbox{$I \subset \left\{0,\ldots,D-1 \right\}$} of relevant coordinates and consider the following truncation:
\begin{equation}
 |\psi_I \rangle = \frac{P_I |\psi \rangle}{\|P_I |\psi \rangle\|_{\ell_2}} \quad \text{where} \quad P_I = \sum_{i \in I} |i \rangle \! \langle i|.
 \label{eq:truncation} 
\end{equation}
This procedure zeros out every amplitude that does \emph{not} belong to the set $I$ and subsequently re-scales the vector to have unit length again. 
Hence, it has the same effect as eliminating nodes in a decision diagram (with subsequent re-scaling) if \(I\) is chosen appropriately.
The following mathematical statement asserts that that fidelity behaves nicely under such truncation procedures.

\begin{lemma} \label{lem:main}
Fix two quantum states $|\psi \rangle, |\phi \rangle$ and a truncation procedure \eqref{eq:truncation}. Then,
\begin{equation*}
F (|\psi \rangle, |\phi_I \rangle ) = F (|\psi \rangle, |\psi_I \rangle) \cdot F (|\psi_I \rangle, |\phi_I \rangle).
\end{equation*}
\end{lemma}
To spell it out, the fidelity between a target state $|\psi \rangle$ and an approximation of $|\phi \rangle$ factorizes into the fidelity of loss incurred by approximating the target times the fidelity between both approximations. 
\begingroup
\setlength\abovedisplayskip{3pt}
\setlength\belowdisplayskip{3pt}
\begin{proof}
The statement readily follows from combining the following properties of a truncation procedure \eqref{eq:truncation}:
\begin{equation*}
P_I |\psi_I \rangle =  |\psi_I \rangle
\quad \text{and} \quad \|P_I |\psi \rangle \|_{\ell_2} = \sqrt{F \left( |\psi \rangle, |\psi_I \rangle \right)}.
\end{equation*}
The first equation states that projecting twice does not change the state, while the second relation expresses the scaling parameter in terms of fidelity. 
Combining both yields
\begin{align*}
F \left( |\psi \rangle, |\phi_I \rangle \right)&= |\langle \psi| P_I |\phi_I \rangle |^2 \\
	&= | \sqrt{ F(|\psi \rangle, |\psi_I \rangle)} \cdot  \langle \psi_I |\phi_I \rangle |^2 \\
	&= F(|\psi \rangle, |\psi_I \rangle ) \cdot F (|\psi_I \rangle, |\phi_I \rangle ).
\end{align*}
\end{proof}

This observation addresses the (possibly) multiple approximation rounds in our simulation approach with a starting state~\(\ket{\chi}\) and unitaries~\(U_i\):
\begin{align*}
|o \rangle &= U_3 U_2 U_1 |\chi \rangle & \text{(true target)}, \\
|o' \rangle &= U_3 | (U_2 U_1 \chi)_I \rangle & \text{(one approximation)}, \\
|o'' \rangle &= U_3 | (U_2 (U_1 \chi)_J)_I \rangle & \text{(two approximations)}.
\end{align*}
Unitary invariance of the fidelity function will allow us to ignore $U_3$. Set
\begin{align*}
|\psi \rangle &= U_3^\dagger |o \rangle = U_2 U_1 |\chi \rangle,&
|\psi_I \rangle &= |(U_2 U_1 \chi)_I \rangle = U_3^\dagger |o' \rangle, \\
|\phi \rangle &= U_2 |(U_1 \chi)_J \rangle,&
|\phi_I \rangle &= |(U_2 (U_1 \chi)_J)_I \rangle = U_3^\dagger |o'' \rangle
\end{align*}
and observe
\begin{align*}
F (|o \rangle, |o'' \rangle) &= F(U_3 |\psi \rangle, U_3|\phi_I \rangle) = F (|\psi \rangle, |\phi_I \rangle ) \\
&= F (|\psi \rangle, |\psi_I \rangle) \cdot F (|\psi_I \rangle, |\phi_I \rangle) \\
&= F (U_3^\dagger |o \rangle, U_3^\dagger |o' \rangle ) \cdot F (U_3^\dagger |o' \rangle, U_3^\dagger |o'' \rangle) \\
&= F (|o \rangle, |o' \rangle ) \cdot F (|o' \rangle, |o'' \rangle ).
\end{align*}
\endgroup

Hence, we have shown that the overall fidelity after multiple approximation rounds can be calculated by simply multiplying the fidelities between the individual rounds.

\begin{table*}[!t]
	\centering
	\caption{Results of the Empirical Validation}
	\label{tab:results}
	\vspace{-1em}
	\scalebox{1}[0.95]{\begin{tabular}{llr@{\hskip 2em}rr@{\hskip 2em}rrlrr}
		\toprule
		Approach & \multicolumn{2}{c}{Benchmark} & \multicolumn{2}{c}{Non-Approximating} & \multicolumn{5}{c}{Proposed Approach} \\
		\cmidrule(r{1em}){2-3}\cmidrule(r{1em}){4-5}\cmidrule{6-10}
		&Name & Qubits                 & Max. DD Size & Runtime [s]                & Max. DD Size & Rounds & \(f_\text{round}\) & Runtime [s] & \(f_\text{final}\) \\
		\midrule
		Memory-driven
		&qsup\_4x5\_15\_0 & 20         & 2\,097\,150 & 3\,666.87         & 1\,963\,906 & 91&0.99 & 5\,512.16 & 0.401  \\
		&&&&                                                                  & 1\,810\,948 & 90&0.975 & 3\,340.89 & 0.102  \\
		&&&&                                                                  & 1\,471\,425 & 89&0.95 & 1\,341.53 & 0.010  \\
		&qsup\_4x5\_15\_1 & 20         & 2\,097\,150 & 2\,024.83         & 1\,417\,398 & 84&0.99 & 1\,066.64 & 0.430  \\
		&&&&                                                                  &    932\,915 & 84&0.975 &    697.40 & 0.119  \\
		&&&&                                                                  &    799\,830 & 49&0.95 &    361.01 & 0.081  \\
		&qsup\_4x5\_15\_2 & 20         & 2\,097\,150 & 2\,090.09         & 1\,963\,347 & 81&0.99 & 3\,208.59 & 0.443  \\
		&&&&                                                                  & 1\,823\,513 & 83&0.975 & 2\,349.31 & 0.122  \\
		&&&&                                                                  & 1\,562\,367 & 84&0.95 & 1\,227.10 & 0.013  \\[3pt]
		Fidelity-driven 
		& shor\_33\_5   & 18             & 73\,736         & 0.50                      &   8\,135            & 6&0.9 &  0.33        & 0.567 \\
		(target \SI{50}{\percent})
		&shor\_55\_2   & 18              & 131\,254        & 0.57                      &   5\,637          & 6&0.9 &  0.20        & 0.559 \\
		&shor\_69\_2   & 21              & 523\,410        & 8.50                      &  52\,726          & 4&0.9 &  1.87        & 0.661 \\
		&shor\_221\_4  & 24              & 1\,472\,942     & 12.56                     &   7\,647          & 5&0.9 &  0.19        & 0.616 \\ 
		&shor\_323\_8  & 27              & 11\,829\,160    & 807.52                    &  13\,706          & 6&0.9 &  0.79        & 0.571 \\ 
		&shor\_629\_8  & 30              & --              & \emph{Timeout}            &  57\,710          & 5&0.9 &  2.07        & 0.596 \\ 
		&shor\_1157\_8 & 33              & --              & \emph{Timeout}            & 535\,001          & 5&0.9 & 117.19       & 0.610 
		\\
		\bottomrule 
	\end{tabular}}
	
	\vspace{3pt}
	\raggedright The runtime \emph{Timeout} indicates the experiment was terminated after \SI{3}{\hour}.
	\vspace{-1.75em}
\end{table*}

\section{Empirical Validation}
\label{sec:evaluation}

To validate the impact of approximation in DD-based quantum circuit simulation, we implemented both methods proposed in Section~\ref{sec:approximation-simulation} on top the simulator provided at \url{https://iic.jku.at/eda/research/quantum_simulation}~\cite{DBLP:journals/tcad/ZulehnerW19,DBLP:conf/iccad/ZulehnerHW19} as part of the JKQ~toolset~\cite{wille2020jkq}.
We used the simulation without approximation as a reference to quantify the effects of approximations with respect to maximum DD size and runtime when approximately simulating the exact circuits.
The simulations were performed on a server running GNU/Linux with a \SI{4.2}{\giga\hertz} (8~cores) CPU and \SI{32}{\gibi\byte} main memory. 
GCC~7.5.0 served as compiler and GNU~parallel~\cite{Tange2011a} was used to control experiment execution.

Table~\ref{tab:results} shows the validation results for the memory- as well as the fidelity-driven approximate simulation method.
The results show that suitably chosen hyper-parameters enable significant improvements for quantum-supremacy circuits from Google and speed-ups of several orders of magnitudes for Shor's algorithm.

The memory-driven approach was validated using the quantum-supremacy circuits from Google using conditional phase-gates~\cite{boixo2016characterizing} (denoted \enquote{qsup\_\(A\)x\(B\)\_\(C\)} with \(A\times B\) being the grid size and \(C\) being the depth).
Since these circuits are designed to be hard to simulate (exactly and approximately) and have little to no redundancy, they are exceptionally challenging for decision diagrams.
Nonetheless, the approximation scheme with a sensible threshold halves the runtime in the best case, while maintaining a fidelity above \SI{10}{\percent}, which (depending on the research context) is still acceptable and, in fact, better than the results from a physical quantum computer~\cite{arute2019quantum,markov2020massively}.
However, the validation also highlights that the parameters have to be carefully selected or there is risk of performance degradation. 

The fidelity-driven approach is ideal if the user has knowledge of the required accuracy as it enables to guarantee a lower bound.
In the validation, we focused on Shor's algorithm~\cite{zulehner2019matrix} (denoted \enquote{shor\_\(A\)\_\(B\)} with the number to factorize \(A\) and coprime \(B\)).
Using the proposed approximate simulation, we were able to simulate Shor's algorithm with a speed-up of several orders of magnitude while setting \SI{50}{\percent} as minimum fidelity.
Notably, the benchmark shor\_1157\_8 (33 qubits) timed out in three hours, while the approximate simulation completed in just under two minutes.
In the experiment, we exploited the knowledge that the inverse \emph{Quantum Fourier Transformation} (QFT) at the end of the algorithm required by far the most time to simulate and, hence, applied the approximation rounds during the inverse QFT.
While \SI{50}{\percent} fidelity seems low, we were able to correctly factorize the numbers given in the benchmarks by performing the non-quantum postprocessing steps of Shor's algorithm.

Our results clearly demonstrate the potential of DD-based simulation with approximation: it enables users to improve runtime performance by up to several orders of magnitude, while keeping the decrease in the accuracy of the resulting quantum state at a level which still delivers meaningful results.

\section{Conclusions}
\label{sec:conclusions}

Quantum circuit simulation is an important pillar of design automation for quantum computing.
In this work, we proposed two new methods for simulation that facilitate an increased performance in a tradeoff with the accuracy of the resulting state.
More precisely, we have investigated quantum circuit simulation with multiple approximation rounds to reduce the size of the decision diagram representing the quantum state.
The first method focuses on efficiency and reactively sacrifices accuracy for a more compact representation.
In the second method, given a minimal required accuracy, the approximation rounds are proactively configured to guarantee the required accuracy.
We further provided an analytical proof for the intuition that the end-to-end fidelity after multiple approximation rounds is the product of the fidelity for each round.
The empirical validation of both methods showed significant increases in performance up to orders of magnitude in the best-case scenario, while still producing a result with suitable accuracy.

\section*{Acknowledgments}
This work has partially been supported by the LIT Secure and Correct Systems Lab funded by the State of Upper Austria as well as by BMK, BMDW, and the State of Upper Austria in the frame of the COMET Programme managed by FFG.

\begingroup
\small

\let\oldbibliography\thebibliography
\renewcommand{\thebibliography}[1]{%
  \oldbibliography{#1}%
  \setlength{\itemsep}{2pt plus 2pt minus 2pt}%
}
\bibliographystyle{abbrv}
\bibliography{lit_header,lit_references}
\eject
\endgroup
\end{document}